\documentstyle[preprint,aps]{revtex}

\def\s{\sigma}

\def\lll{\lambda}

\newcommand{\be}{\begin{equation}}
\newcommand{\ee}{\end{equation}}

\input epsf

\begin{document}
\preprint{WISC-MILW-96-TH-10}
\title{Validity of the Semiclassical Approximation \\
and Back-Reaction}
\author{Sukanta Bose,\footnote{Electronic address: {\em bose@csd.uwm.edu}}  
Leonard Parker,\footnote{Electronic address: {\em leonard@cosmos.phys.uwm.edu}} 
and Yoav Peleg\footnote{Electronic address: {\em yoav@csd.uwm.edu}} }
\address{Department of Physics \\
University of Wisconsin-Milwaukee, P.O.Box 413 \\
Milwaukee, Wisconsin 53201, USA}
\maketitle
\vskip 0.5cm
\begin{abstract}
Studying two-dimensional evaporating dilatonic 
black holes, we show that the semiclassical approximation, 
based on the background field approach, is valid 
everywhere in regions of weak curvature (including the horizon), 
as long as one takes into account 
the effects of back-reaction of the Hawking radiation on
the background geometry. 
\end{abstract}

\newpage 

The suggestion that quantum gravitational effect may play a role 
in the evaporation of black holes \cite{tHooft} received some attention 
recently \cite{Mikovic1}. 
Since we do not have a full theory of quantum gravity it is not clear 
how to define a criterion according to which we can check 
the validity of the semiclassical approximation. 't Hooft suggests 
that the resolution of the information puzzle (in a unitary framework) 
is one of the conditions that a quantum theory of gravity should 
satisfy, and if the present semiclassical theory implies a non-unitary 
evolution, it cannot be consistent with the full theory of quantum 
gravity. This may very well be the case, but some try to go 
even further and show that the standard semiclassical approximation 
(based on the background field approach) breaks down in regions of 
weak curvature, just outside the horizon of a black hole 
\cite{Mathur,Gilad}. 
Though in Ref. \cite{Mathur} quantum matter fields forming a black hole 
(and consequently emitting Hawking radiation) 
was considered, the back-reaction of the quantum radiation 
on the geometry was ignored. We show in this work that when we  
take into account the back-reaction, we avoid the conclusions of 
\cite{Mathur}. Namely, we show in the framework of two-dimensional (2D) 
dilaton Gravity (in which we can solve the semiclassical equations 
including the back-reaction) that the semiclassical approximation 
is valid as long as one includes the back-reaction. We therefore conclude 
that, at least in 2D dilaton Gravity,  
the breakdown of the background field method in regions of weak 
curvature is an artifact of the neglection of the back-reaction.  

In this work we use the same criterion as in \cite{Mathur} 
according to which we check the validity of the semiclassical 
approximation. One considers two almost identical classical 
space-times with a classical dilaton field, which may differ only 
at the Planck scale. Then the evolution of a quantum scalar (matter) field 
on these two classical backgrounds is considered. By taking identical 
space-like hypersurfaces (i.e., hypersurfaces with the same internal 
geometry\footnote{For the hypersurfaces that we consider the shift 
vector does not necessarily vanish at spatial infinity, and one 
can have different ADM masses for the same intrinsic geometry.}) 
in both space-times, one can compare the evolution of 
the two quantum states in the two different space-times. 
For example 
one can take two identical hypersurfaces at asymptotic past 
infinity, and start with a vacuum state on both hypersurfaces. 
Since the initial hypersurfaces are identical one can calculate the 
scalar product of the two states (using the Klein-Gordon scalar 
product defined on the hypersurface). In the collapsing black hole 
geometry one finds that the scalar product of the two vacua is one. 
Namely, the two past asymptotic vacua are the same. 
Then (using the Schr\"{o}dinger picture) one evolves the two states, 
each in its space-time, to late-time identical hypersurfaces (in the 
two different space-times). It was shown in \cite{Mathur} 
that (neglecting the back-reaction) for very special late-time space-like 
hypersurfaces, called the S-hypersurfaces, 
(ones that intersect the infalling matter near the horizon and also 
capture a significant amount of Hawking radiation), 
the scalar product between the two states is almost zero, 
even for spacetimes with a mass difference of the order 
of $\Delta M \sim \exp(-M/M_{P})$, where $M_{P}$ is the Planck mass, 
or its 2D analog. 
Since in this case $\Delta M << M_P$, they conclude that the semiclassical 
approximation is not valid. 
We are going to show in this work that even for the special hypersurfaces 
considered in \cite{Mathur}, if we include the back-reaction, the 
scalar product is not zero, but almost one even for $\Delta M \sim M_P$. 
This suggests that the semiclassical approximation is valid in 
regions of weak curvature as long as we include the back-reaction. 

We study a modified theory of 2D dilaton gravity, 
described by the action \cite{BPP} 
  \begin{eqnarray} \label{action}
S_{\mbox{mod}} &=& {1\over 2\pi}\int d^2x \sqrt{-g(x)} 
\left[ ( e^{-2\phi} - \kappa \phi) R(x) \right. \\
&+& (4 e^{-2\phi} + \kappa) (\nabla \phi)^2 + 
4 \lambda^2 e^{-2\phi} - {1\over 2} \sum_{i=1}^{N} 
\left. (\nabla f_i)^2 \right] \nonumber \\
&-& {\kappa \over 8\pi} \int d^2x \sqrt{-g(x)} \int d^2x' \sqrt{-g(x')} 
R(x) G(x,x') R(x') , \nonumber
  \end{eqnarray}
where $R(x)$ is the 2D Ricci scalar, $\phi$ is the dilaton field, 
$f_i$ are $N$ massless scalar fields, $\kappa = N \hbar / 12$, and 
$G(x,x')$ is an appropriate Green's function for $\nabla^2$. 
The effective action (\ref{action}) describes the full quantum 
theory in the large $N$ limit, in which case the fluctuations of $\phi$ 
and $g_{\mu \nu}$ can be neglected. Namely, the semiclassical 
approximation is exact at the large $N$ limit. 
For finite $N$, the action (\ref{action}) is a semiclassical 
approximation to the full quantum theory. Nevertheless it includes the 
back-reaction of the quantum scalar fields on the classical 
metric and dilaton fields. In this work we do not take the large $N$ 
limit, and so the action is only a semiclassical effective action, 
and we assume that it is a good approximation as long as the 
curvature and the coupling are small. 

One can derive the effective action (\ref{action}) by first fixing 
the diffeomorphism gauge and then quantizing the reduced system 
\cite{Mikovic1}. After choosing an appropriate initial quantum state, 
one gets the action (\ref{action}) as the semiclassical effective 
action \cite{Mikovic2}. 

It is convenient to use null coordinates and conformal gauge, 
$g_{+-}=-e^{2 \rho}/2$, and $g_{++}=g_{--}=0$. In Kruskal coordinates, 
$x^{\pm}$, for which $\phi(x^+,x^-)=\rho(x^+,x^-)$, the evaporating 
black hole solution of (\ref{action}) is \cite{BPP} 
  \be \label{blackholecomplete}
e^{-2\phi} = e^{-2\rho} = -\lambda^2 x^+ x^- - {M\over \lambda x^+_0} 
(x^+ - x^+_0) \Theta (x^+ - x^+_0) - {\kappa\over 4} 
\ln(-\lambda^2 x^+ x^-) . 
  \ee
This solution describes the formation and subsequent evaporation of a 
black hole which forms by an infalling shock wave \cite{BPP} 
with mass $M$. In Fig. 1 we show its Penrose diagram. 

Before the shock wave the solution is a static vacuum solution 
with no Hawking radiation, 
  \be \label{staticsolution}
{\left( e^{-2\phi} \right)}_{x^+ < x^+_0} = -\lll^2 x^+ x^- - 
{\kappa\over 4} \ln(-\lll^2 x^+ x^- ) = \exp(2\lll y) -{\kappa \lll 
y\over 2} \; , 
  \ee 
where $y = (y^+ - y^-)/2$, and $ \lll y^{\pm} = \pm \ln(\pm \lll x^{\pm})$ 
are the asymptotically flat coordinates in the static region 
$x^+ < x^+_0$. The solution (\ref{staticsolution}) includes a region 
of strong coupling $e^{2\phi} >> 1$. In order to avoid 
that region in which the semiclassical approximation is not valid 
we impose reflection boundary condition on a timelike hypersurface 
$y = y_b =\mbox{const}$, such that $\exp[2\phi(y_b)] << 1$,   
and consider the solution only in the weak coupling region $y > y_b$. 

After the shock wave we have an evaporating black hole space-time, 
  \be \label{blackhole}
{\left( e^{-2\phi} \right)}_{x^+ > x^+_0} = -\lll^2 x^+ \left( x^- + 
{M\over \lll^3 x^+_0} \right) - {\kappa\over 4} \ln(-\lll^2 x^+ x^-) 
+ {M\over \lll}. 
  \ee 
In this region the asymptotically flat coordinates are 
$\lll \sigma^+ = \ln(\lll x^+)$ and $\lll \sigma^- = 
- \ln\{ -[\lll x^- + M/(\lll^2 x^+_0)] \} $. The black hole singularity 
is initially hidden behind a timelike apparent horizon. As the 
black hole evaporates by emitting Hawking radiation, the singularity 
meets the shrinking horizon in finite retarded time to become naked, 
at $x^- = x^-_{int}$, see Fig. 1.  

We denote the spacetime (\ref{blackholecomplete}) by ${\cal M}$. 
One can consider another spacetime described by 
(\ref{blackholecomplete}) but with a different mass, $\bar{M} = 
M + \Delta M$. Since the Kruskal coordinates for which $\phi = \rho$ 
depend on the solution and therefore on the ADM mass of the 
spacetime, the Kruskal coordinates for $\bar{M}$ 
will be denoted by $\bar{x}^{\pm}$. We denote the spacetime 
(\ref{blackholecomplete}) with $\bar{M}$ and $\bar{x}^{\pm}$ by   
$\bar{\cal M}$. In the region $x^+ < x^+_0$ ($\bar{x}^+ < x^+_0$),  
the space-times ${\cal M}$ and 
$\bar{\cal M}$ are the same, so we are interested in the 
region $x^+ > x^+_0$. In this region we have evaporating black 
holes, one with mass $M$ in ${\cal M}$ and the other with mass 
$\bar{M}$ in $\bar{\cal M}$. Without loss of generality, we 
take $\lll x^+_0 = 1$. A necessary condition for the validity of the 
semiclassical approximation is that $M >> \kappa \lll$ ($\kappa \lll$ 
may be considered as the 2D analog of the Planck mass, $M_P$ 
\cite{RST}).  

The last term on the right-hand-side of Eq. (\ref{blackholecomplete}) is 
due to the back reaction. If we set $\kappa = 0$ in 
(\ref{blackholecomplete}) we get  
  \be \label{CGHSsolution}
e^{-2\phi} = e^{-2\rho} = -\lambda^2 x^+ x^- - {M\over \lambda x^+_0} 
(x^+ - x^+_0) \Theta (x^+ - x^+_0) , 
  \ee
which is exactly the classical CGHS solution \cite{CGHS}, 
that was considered in Ref. \cite{Mathur}. 
The CGHS solution (\ref{CGHSsolution}) 
is the one with no back reaction, and so taking 
the limit $\kappa \rightarrow 0$ in our semiclassical solution 
is equivalent to ignoring the back-reaction. We would like to 
stress that the solution (\ref{CGHSsolution}) does not describe a 
non-unitary evolution, because the black hole in (\ref{CGHSsolution}) 
does not evaporate away. It is the solution (\ref{blackholecomplete}) that
may describe a non-unitary evolution \cite{BPP}, and according to
't Hooft one should examine the validity of the semiclassical 
approximation that lead to the solution (\ref{blackholecomplete}) 
and not (\ref{CGHSsolution}).  

We follow \cite{Mathur} and calculate the scalar product of two 
initial vacuum states in ${\cal M}$ and $\bar{\cal M}$. 
First we calculate it on past asymptotic hypersurfaces 
and then on future S-hypersurfaces. We do the calculations for a finite 
$\kappa$. 
When $\kappa = 0$ we recover the results of \cite{Mathur}. 
As we noted above, this limit corresponds to 
ignoring the back-reaction which is inconsistent with the quantum 
Hawking radiation. By expressing the results 
explicitly in terms of $\kappa$, we find that the scalar product 
is almost one for any Planck scale perturbations. 

We define the space-like hypersurfaces, $\Sigma$, 
(which are one dimensional in our 2D
theory) by means of their intrinsic geometry, 
which can be determined by giving the dilaton field $\phi$, and its 
first derivative $d\phi/ds$, where $s$ is the proper distance 
along the space-like hypersurface \cite{Mathur}. 
The two different semiclassical 
solutions are described by (\ref{blackhole}) with different masses, 
$M$ and $\bar{M} \equiv M + \Delta M$. Let $\bar{\Sigma}$ be the spacelike 
hypersurface in $\bar{\cal M}$ having the same intrinsic geometry as 
$\Sigma$ in ${\cal M}$. We denote this equivalence 
relation by $\Sigma = \bar{\Sigma}$. 
The 1D hypersurfaces $\Sigma$ and 
$\bar{\Sigma}$ can be written in the form 
  \begin{eqnarray}
x^- &=& x^-_{\Sigma}(x^+)   \qquad , \qquad \mbox{$\Sigma$-hypersurface
in ${\cal M}$} 
\label{sigma} \\
\bar{x}^- &=& \bar{x}^-_{\bar{\Sigma}}(\bar{x}^+) \qquad , \qquad \mbox{
$\bar{\Sigma}$-hypersurface in $\bar{\cal M}$ .} \label{sigmabar}
  \end{eqnarray} 
Let us define 
  \be \label{fdef}
f(\lambda x^+) \equiv \lambda x^-_{\Sigma}(x^+) + {M\over \lll} \qquad 
\mbox{and} \qquad \bar{f}(\lambda \bar{x}^+) \equiv \lll 
\bar{x}^-_{\bar{\Sigma}}(\bar{x}^+) + {\bar{M}\over \lll} 
  \ee
Then using (\ref{blackhole}) and (\ref{fdef}), the equations 
$\phi(x^+,x^-_{\Sigma})=\phi(\bar{x}^+,\bar{x}^-_{\bar{\Sigma}})$, 
and $d \phi(x^+,x^-_{\Sigma}) / ds =
d \phi(\bar{x}^+,\bar{x}^-_{\bar{\Sigma}}) / d\bar{s}$, 
(which imply that $\Sigma = \bar{\Sigma}$), become 
  \be
{M\over \lll} - \lll x^+ f - {\kappa\over 4} \ln[\lll x^+ 
(f - M/\lll)] = {\bar{M}\over \lll} - \lll \bar{x}^+ \bar{f} 
- {\kappa\over 4} \ln[\lll \bar{x}^+ ( \bar{f} - \bar{M}/
\lll)] \label{phiequation} 
  \ee
  \be
{ f + \lll x^+ f' + {\kappa\over 4} \left( 
{1\over \lll x^+} + {f'\over f - M/\lll} \right) \over 
\sqrt{ - f'} } =  { \bar{f} + \lll \bar{x}^+ {\bar{f}}' 
+ {\kappa\over 4} \left( {1\over \lll \bar{x}^+} + {{\bar{f}}'
\over \bar{f} - \bar{M}/\lll} \right) \over 
\sqrt{ - \bar{f}' } }\label{dphiequation} , 
  \ee
where $f = f(\lll x^+)$, $\bar{f} = \bar{f}(\lll \bar{x}^+)$, and 
$'$ denotes derivative with respect to the argument of the function. 
Eqs. (\ref{phiequation}) and (\ref{dphiequation}) reduced to the 
corresponding expressions in \cite{Mathur} when the back-reaction 
is ignored, (i.e., when $\kappa = 0$). We must find 
the solutions of (\ref{phiequation}) and 
(\ref{dphiequation}), $\bar{x}^+ = 
\bar{x}^+(x^+)$ and $\bar{x}^- = \bar{x}^-(x^-)$, in order 
to determine the scalar product of the two vacua.  

Consider first past asymptotic space-like hypersurfaces (like the   
$\Sigma^0$-hypersurface in Fig. 1) in ${\cal M}$ or 
$\bar{\cal M}$. A past asymptotic hypersurface in ${\cal M}$ 
can be given by the equation 
  \be \label{sigma0}
f(\lll x^+) = - A^2 \lll x^+  \qquad \qquad \mbox{the $\Sigma^0$ 
hypersurface},  
  \ee
where $A^2$ is a constant satisfying 
$A^2 >> M/\lll$ in order for the $\Sigma^0$-hypersurface to 
cross the infalling null matter long before the apparent horizon 
is formed. The hypersurface (\ref{sigma0}) corresponds to a fixed 
asymptotic time, namely, $t = (\sigma^+ + \sigma^-)/2 = 
- (2 \lll)^{-1} \ln [ - f(\lll x^+)/ x^+ ] = - \lll^{-1} \ln(|A|) 
= \mbox{const}$. 
For the $\Sigma^0$ hypersurface (\ref{sigma0}) one can show that in 
Eqs. (\ref{phiequation}) and (\ref{dphiequation}) $\kappa$ appears only 
in the combination $\kappa / A^2$. Since $A^2 >> M/\lll$ and 
$M/\lll >> \kappa$, one can neglect terms of order $\kappa/A^2$. 
Namely, the effect of back-reaction is negligible on the $\Sigma^0$
hypersurface. This is expected since the Hawking radiation (and so 
the back-reaction) starts when the apparent horizon is formed, 
to the far future of $\Sigma^0$.  
Neglecting $\kappa / A^2$ one can easily solve 
(\ref{phiequation}) and (\ref{dphiequation}) exactly (see \cite{Mathur}) 
to get 
  \be \label{sigma0xpm}
\lll \bar{x}^+ = \lll x^+ + \sqrt{{\Delta M\over \lll A^2}} 
\qquad , \qquad \lll \bar{x}^- = \lll x^- + 
\sqrt{{A^2 \Delta M\over \lll}} - {\Delta M\over \lll}. 
  \ee
We see that as $\Sigma^0$ approaches $\Im^-$, i.e., as 
${M/\lll \over A^2} \rightarrow 0$, we have 
$\bar{x}^+ = x^+$. So on $\Im^-$ and $\bar{\Im}^-$ the vacua 
that are defined with respect to the modes $\exp(-i k_+ \sigma^+)$ 
and $\exp(-i \bar{k}_+ \bar{\sigma}^+)$ are the same, because 
$\bar{\sigma}^+ = \lll^{-1} \ln(\lll \bar{x}^+) = \lll^{-1} \ln(
\lll x^+) = \sigma^+$. 

The strategy is therefore to consider a vacuum state on $\Im^-$ and 
$\bar{\Im}^-$, to evolve it to a future hypersurface, $\Sigma^f$, once 
embedded in ${\cal M}$ and then embedded in $\bar{\cal M}$. In ${\cal M}$ 
the vacuum state corresponds to the modes $\exp[-i k_+ \ln(\lll x^+)/\lll]$, 
while in $\bar{\cal M}$ it corresponds to the modes 
$\exp[-i \bar{k}_+ \ln(\lll \bar{x}^+)/\lll]$. In order to calculate the 
scalar product on $\Sigma^f$, we need to know $\bar{x}^+$ as a function 
of $x^+$ on $\Sigma^f$. Unlike on $\Sigma^0$, on $\Sigma^f$ we do not 
necessarily have $\bar{x}^+ = x^+$. We have to solve (\ref{phiequation}) 
and (\ref{dphiequation}) on $\Sigma^f$ and find $\bar{x}^+ = \bar{x}^+(x^+)$. 

The future space-like hypersurfaces that we consider are the 
S-hypersurfaces described in \cite{Giddings,Mathur}. Those can be 
written in ${\cal M}$ as
  \be \label{shypersurface}
f(\lll x^+) = - \alpha^2 \lll x^+ - 2 \alpha \sqrt{{M\over \lll}}
\qquad \qquad \mbox{the S-hypersurfaces} , 
  \ee
where $\alpha$ is a very small constant. An example for a S-hypersurface 
is shown in Fig. 1. Let $\gamma M$ be the amount 
of Hawking radiation that is captured by the S-hypersurface, where 
$\gamma$ is a constant between one and zero, then one 
can use (\ref{blackhole}) to show that $\alpha$ is of the order of 
  \be \label{orderalpha}
\alpha \sim \sqrt{{M\over \lll}} \exp\left( - {4 \gamma M \over  
\kappa \lll} \right).  
  \ee
Since $M/\kappa \lll >> 1$ we see that $\alpha$ is 
exponentially small (i.e., non-perturbatively small in terms of 
$\kappa \lll/M$). 

Since by construction 
for $M = \bar{M}$ we have $\bar{x}^{\pm} = x^{\pm}$, we expand 
$\Delta x^{\pm}=\bar{x}^{\pm} - x^{\pm}$ 
in powers of $\Delta M / M$. We assume that the 
leading term is of order $\Delta M / M$, which will turn out to be 
self-consistent. We therefore write  
  \be 
\lll \bar{x}^+ = \lll (x^+ + \Delta x^+) 
\equiv \lll x^+ + {\cal E}^+(\lll x^+) {\Delta M\over M} 
+ {\cal O}{\left( \Delta M \over M \right)}^2 \label{deltaplus} 
  \ee
  \be
\lll \bar{x}^- = \lll(x^- + \Delta x^-) 
\equiv \lll x^- + {\cal E}^-(\lll x^-) {\Delta M\over M} 
+ {\cal O}{\left( \Delta M \over M \right)}^2 \label{deltaminus} ,
  \ee
where ${\cal E}^{\pm}$ are functions to be determined. 
We assume that for $\Delta M / M << 1$ the term of order 
$(\Delta M/M)^2$ can be neglected. 

Using (\ref{deltaplus}), (\ref{deltaminus}) and (\ref{shypersurface}), 
Eqs. (\ref{phiequation}) and (\ref{dphiequation}) become
  \begin{eqnarray} \label{phis}
\left( \lll x^+ - {\kappa \over 4 (\alpha^2 \lll x^+ + 2\alpha 
\sqrt{M/\lll} + M/\lll) } \right) {\cal E}^- &=& \\
= \left( \alpha^2 \lll x^+ + 2\alpha \sqrt{{M\over \lll}} - 
{\kappa \over 4\lll x^+} \right) {\cal E}^+ 
&-& (\lll x^+ - 1) {M\over \lll}
\nonumber
  \end{eqnarray}
and 
  \begin{eqnarray} \label{dphis}
& & \left( - \alpha \sqrt{{M\over \lll}} + 
{\kappa\over 8 \lll x^+} - {\kappa \alpha^2 \over 
8 (\alpha^2 \lll x^+ + 2\alpha \sqrt{M/\lll} + M/\lll)} 
\right) \left[ ({\cal E}^+)' - ({\cal E}^-)' \right] = 
\nonumber \\
& & = \left( {\kappa \over 4 (\lll x^+)^2} + \alpha^2 \right) {\cal E}^+ 
- \left( 1 + {\kappa \alpha^2 \over 4 (\alpha^2 \lll x^+ + 2\alpha 
\sqrt{M/\lll} + M/\lll)^2 } \right) {\cal E}^- + {M\over \lll}\; . 
  \end{eqnarray}
Since $\alpha$ and $\kappa$ are both small compare to one, we solve 
(\ref{phis}) and (\ref{dphis}) by expanding in powers of $\kappa$ and 
$\alpha$. The solutions are 
  \be
{\cal E}^+ = {- M/\lll \over 2\alpha \sqrt{M/\lll} - \kappa/(4 \lll x^+)} 
\left( 1 - {\lll \kappa \over 4 M} + {\cal O}(\kappa^2,\alpha^2) \right) 
\label{ep} 
  \ee
  \be
{\cal E}^- = - {M\over \lll} \left( 1 + {\alpha^2 \over 2\alpha 
\sqrt{M/\lll} - \kappa/(4\lll x^+)} + {\cal O}(\kappa^2,\alpha^2) 
\right) \; . 
\label{em}
  \ee
Setting $\kappa = 0$ gives
  \be \label{nobackreaction}
({\cal E}^+)_{\kappa=0} = - {\sqrt{M/ \lll} \over 2\alpha} 
\quad , \quad ({\cal E}^-)_{\kappa=0} = - {M\over \lll}\left( 1 + 
{\alpha \over 2 \sqrt{M/\lll}} \right)  
\quad \qquad \mbox{ignoring back-reaction} ,
  \ee
which are exactly the results of \cite{Mathur}, as expected. 
Since $\alpha$ is exponentially small (see (\ref{orderalpha})), the shift in 
$x^+$, i.e., ${\cal E}^+$, is very large in (\ref{nobackreaction}). 
This leads to a breakdown of the 
semiclassical approximation. However one should remember that in order 
to get (\ref{nobackreaction}) we assume 
that $\kappa$ is much smaller than $\alpha \sqrt{M/\lll}$. 
Is this consistent with (\ref{orderalpha})? 
Using $M/(\kappa \lll) >> 1$ and (\ref{orderalpha}) we see that 
  \be \label{smallalpha}
\sqrt{{M\over \lll}} \alpha \sim {M\over \lll} \exp \left( 
- {4 \gamma M \over \kappa \lll} \right) <<< {M\over \lll} 
{\kappa \lll \over M} = \kappa , 
  \ee
where $X >>> Y$ means that $X$ is non-perturbatively larger than $Y$
(i.e., there exist a constant $c < 1$, such that $(Y/X) < c^n$, 
for any $n \in {\cal N}$). 
We see that for the S-hypersurfaces $\kappa$ is non-perturbatively 
{\em larger} than $\alpha \sqrt{M/\lll}$, and therefore we cannot 
take $\kappa = 0$ in (\ref{ep}) and (\ref{em}). 
On the other hand, we definitely can take $\alpha = 0$, and get 
  \be \label{epm0}
({\cal E}^+)_{\alpha=0} = {4M x^+\over \kappa} \left( 1 - 
{\lll \kappa \over 4 M} \right) \quad , \quad 
({\cal E}^-)_{\alpha=0} = - {M \over \lll} \quad 
\qquad \mbox{including back-reaction} ,  
  \ee
or for $\bar{x}^{\pm}$ 
  \be \label{xpm} 
\lll \bar{x}^+ \simeq \lll x^+ \left( 1 + {\Delta M\over \kappa \lll /4} 
- {\Delta M \over M} \right)  
\qquad , \qquad \lll \bar{x}^- \simeq  \lll x^- - {\Delta M \over \lll} .
  \ee 
From (\ref{xpm}) we see that the relations between the asymptotically 
flat coordinates $\bar{\sigma}^{\pm}$ and $\sigma^{\pm}$ are 
approximately linear on the S-hypersurface. In particular, 
  \be \label{sigmarelations}
\bar{\sigma}^+ = \lll^{-1} \ln(\lll \bar{x}^+) \simeq \sigma^+ 
+ \lll^{-1} \ln(1 + 4\Delta M /\kappa \lll) . 
  \ee
It follows that the Bogoliubov coefficients, $\beta_{k,\bar{k}}$, 
obtained from the Klein-Gordon scalar product, vanish: 
$\beta_{k,\bar{k}} \propto
(\exp(i \bar{k}_+ \bar{\s}^+), \exp(-i k_+ \s^+)) = 0$. Hence 
the Fock space scalar product \cite{Umezawa}, 
$\langle 0 | \bar{0} \rangle = (det(1 + \beta^{\dag} 
\beta ))^{-1/2}$, between the vacua defined with respect to the 
modes $\exp(-i k_+ \sigma^+)$ and $\exp(-i \bar{k}_+ \bar{\sigma}^+)$ 
is 1. Our results are valid as long as $\Delta x^{\pm} \leq x^{\pm}$, 
namely, when $\Delta M \sim \kappa \lll$.  
This is what we expect from the semiclassical approximation. 


Since the scalar product of $|0\rangle$ and $|\bar{0} \rangle$ 
is 1 on $\Im^-$ (or $\bar{\Im}^-$), the quantum states $|0\rangle$ and 
$| \bar{0} \rangle$ are indistinguishable on $\Im^-$. The fact that the 
scalar product remains nearly one even on late-time hypersurfaces 
(like the S-hypersurface) implies that the states remain indistinguishable 
throughout their evolution in the different backgrounds ${\cal M}$ and 
$\bar{\cal M}$. This shows that the quantum matter states are insensitive 
to Planck-scale changes in the background geometry. 
Therefore, the semiclassical approximation is valid everywhere in 
regions of weak coupling and curvature including the apparent 
horizon. 

The semiclassical approximation evidently breaks down only near 
the naked singularity, the intersection point in Fig. 1. The 
dilaton field (and the curvature) diverges at that point, and from 
equations (\ref{phiequation}) and (\ref{dphiequation}) 
one expects to get large shifts in the Kruskal 
coordinates, leading to a zero scalar product between the two 
vacua. 

In passing, we note that one can find different initial conditions 
\cite{BPP} in our effective theory 
(\ref{action}) that yield exactly the same solutions as the ones given in 
Eq. (\ref{CGHSsolution}). For those solutions the shifts in $x^{\pm}$ 
are exactly the ones found in \cite{Mathur}, i.e., the ones given 
in Eq. (\ref{nobackreaction}). However, unlike the corresponding example 
considered in \cite{Mathur}, one cannot conclude that the semiclassical 
approximation breaks down for this metric (\ref{CGHSsolution}) in our 
model. This is  because the solutions in 
(\ref{CGHSsolution}) correspond in our effective theory 
to a different vacuum state of the scalar matter fields \cite{BPP}. 
Unlike the collapsing black hole solutions 
(\ref{blackholecomplete}) which correspond to a Schwarzschild 
vacuum, the solutions (\ref{CGHSsolution}) correspond to the Kruskal vacuum. 
Namely, to the vacuum with respect to the modes $\exp(-i k_+ x^+)$. 
The relations between the Kruskal coordinates 
given by (\ref{deltaplus}), (\ref{deltaminus}) and  
(\ref{nobackreaction}) are {\em linear}.  
Hence, the scalar product between the Kruskal vacua is 1, and not 
0. Thus the semiclassical approximation remains valid even for the 
solutions (\ref{CGHSsolution}) in our theory. 

One can also study a larger class of 
exactly solvable 2D semiclassical models 
by considering area-preserving diffeomorphism invariant theories 
\cite{Cruz}, where our model \cite{BPP} arises as a special case. 
Since this larger class of 2D models that include the back-reaction 
can be explicitly solved, one can use our approach to check 
the validity of the semiclassical approximation in this larger class.
On the other hand, 
since we do not have an explicit solution for the four-dimensional 
(4D) evaporating black hole (including the back-reaction), it is not 
straight forward to extend our results to the more realistic 4D 
case. One may try to follow the approach of \cite{Russo} by 
studying an effective 4D evaporating solution and comparing the 
results to those of \cite{Gilad} in which the semiclassical 
approximation is neglected. 
The generalization of our 2D results to 4D is the following conjecture: \\
{\em In four dimensional gravitational collapse to form an 
evaporating black hole, the semiclassical approximation 
is valid everywhere in regions of weak curvature 
(including the apparent horizon), as long as one takes into account 
the back-reaction (including that of the Hawking radiation) on the 
back-ground geometry.}

\vskip 5pt
   
We thank Gilad Lifschytz and Samir Mathur for helpful discussions.  
This work was supported by the National Science Foundation under grant 
PHY 95-07740.

\newpage

\begin{figure}
\caption{Penrose diagram for the evaporating black hole. The lower 
dashed curve is the $\Sigma^0$ hypersurface, and the upper one is 
the S-hypersurface.} 
\label{fig1}
\end{figure}

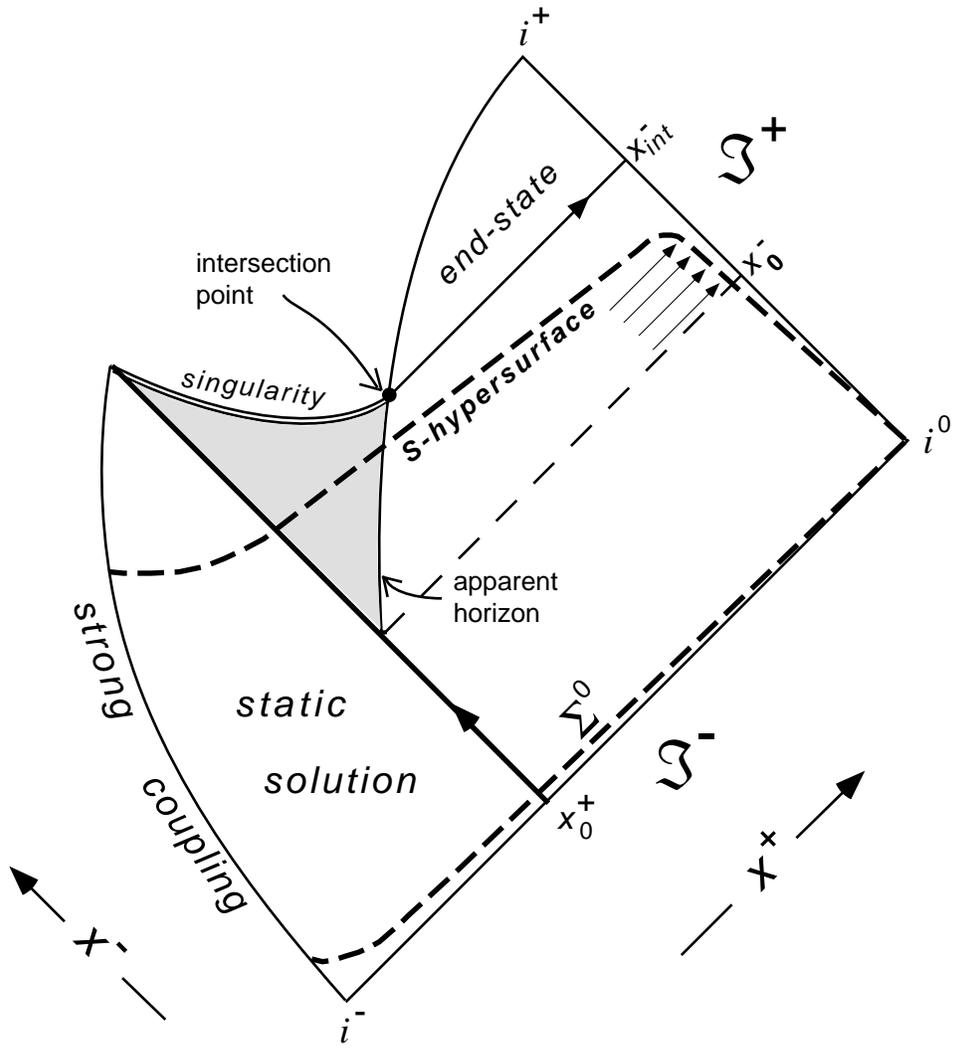
\end{document}